# Coexistence of Antiferromagnetism and Superconductivity in Mn/Nb(110)

Roberto Lo Conte[1,*,§], Maciej Bazarnik[1,2,*,§], Krisztián Palotás[3,4,5], Levente Rózsa[6], László Szunyogh[4,7], André Kubetzka[1], Kirsten von Bergmann[1], Roland Wiesendanger[1]

[1]*Department of Physics, University of Hamburg, D-20355 Hamburg, Germany*
[2]*Institute of Physics, Poznan University of Technology, Piotrowo 3, 60-965 Poznan, Poland*
[3]*Institute for Solid State Physics and Optics, Wigner Research Center for Physics, H-1525 Budapest, Hungary*
[4]*Department of Theoretical Physics, Budapest University of Technology and Economics, H-1111 Budapest, Hungary*
[5]*MTA-SZTE Reaction Kinetics and Surface Chemistry Research Group, University of Szeged, H-6720 Szeged, Hungary*
[6]*Department of Physics, University of Konstanz, D-78457 Konstanz, Germany*
[7]*MTA-BME Condensed Matter Research Group, Budapest University of Technology and Economics, H-1111 Budapest, Hungary*

**We report on the structural and magnetic properties of single and double atomic layers of Mn on a clean and unreconstructed Nb(110) substrate. Low-temperature scanning tunneling spectroscopy measurements reveal a proximity-induced superconducting state in the Mn thin films, which are found to grow pseudomorphically on the Nb surface. Spin-polarized scanning tunneling microscopy measurements reveal a *c*(2x2) antiferromagnetic order in the Mn layers, with an out-of-plane spin-orientation. First-principles density functional theory calculations confirm the experimentally observed magnetic state, which is understood as the consequence of a strong intra- and inter-layer nearest-neighbor antiferromagnetic exchange coupling. These results are expected to be of importance for the design and the investigation of novel superconducting antiferromagnetic spintronic systems.**

The concepts of topological superconductivity and superconducting spintronics sparked recent research efforts in combining magnetic and superconducting materials. It is well understood that when the spin of a magnetic impurity interacts with a superconducting condensate, Yu-Shiba-Rusinov (YSR) in-gap bound states emerge [1–4]. This picture becomes more complex when magnetic adatoms are put together to form one-dimensional atomic chains, where YSR states start hybridizing, giving rise to dispersive energy bands and, potentially, Majorana end modes [5–11]. Analogously, it is expected that when two-dimensional magnetic islands on a superconductor are considered, multiple YSR bands are formed and propagating topological edge modes can emerge [12–14]. Furthermore, spin-triplet pairing amplitudes can exist in magnetic thin films on top of a superconductor [15,16], resulting in spin-polarized supercurrents [17–19]. Spin-supercurrents are particularly interesting from an application point of view, since they can be used for the design of π-type Josephson junctions for superconducting digital and quantum circuits [20,21] as well as for novel ultra-low dissipation spintronic devices such as magnetic domain wall memories [22] and logics [23].

Recently, a successful recipe for the preparation of clean and unreconstructed Nb(110) surfaces was demonstrated [24], making it possible to study the properties of magnetic 3d transition metals impurities [25,26] and atomic chains [10] on the *s*-wave elemental superconductor with the highest critical temperature ($T_C$ = 9.25 K). Interestingly, the study of artificially built Mn atomic chains on top of a clean Nb(110) substrate has shown the emergence of *p*-wave superconductivity [10] and the evidence of interacting Majorana modes [27]. Those results on atomic scale systems motivate the investigation of extended Mn films on top of Nb substrates, where equally interesting emergent effects could arise. However, this would be possible only if the proximity-induced superconductivity is not suppressed by the exchange field of the 2D magnetic system. It has been previously shown that magnetic impurities on a superconducting Pb surface strongly suppress the superconducting phase above a certain concentration level [28]. Accordingly, it is important to understand the feasibility of growing magnetic 3d transition metal films on Nb substrates, where proximity-induced superconductivity and magnetic order coexist. In 1D chains of Mn on Nb(110) the superconductivity survives [10], making Mn an auspicious starting point for thin-film studies. Furthermore, the knowledge of the exact spin texture present in the magnetic layer is key for the understanding of potential emergent electronic properties [29,30], motivating experimental investigations via scanning probe techniques with magnetic sensitivity and atomic resolution.

Here, we report on the study of the structural properties and of the magnetic ground state of superconducting Mn ultrathin films on Nb(110) at 4.2 K. Low-energy scanning tunneling spectroscopy (STS) measurements are carried out to confirm the superconducting state of Mn ultrathin films deposited on a clean Nb(110) surface. Spin-polarized scanning tunneling microscopy (SP-STM) [31] is used for the direct imaging of the morphology and the magnetic ground states of Mn monolayers (MLs) and double-layers (DLs). The experimentally observed magnetic state is successfully reproduced by first-principles density functional theory (DFT) calculations.

The Mn films, 1.2 ML - 1.5 ML in thickness, are grown via physical vapor deposition over a clean and unreconstructed Nb(110) surface, resulting in a complete ML and a partial DL available for experimental investigation. Low-bias STS measurements are performed on a sample where the bare Nb(110) surface as well as Mn-ML and Mn-DL are accessible in the same scanning area (see inset in Fig. 1). The STS point spectrum acquired over the bare Nb substrate (solid cyan curve in Fig. 1) confirms its superconducting state, as shown by the strong reduction in the differential conductance, d$I$/d$U$, around the Fermi level ($U = 0$ mV) and the presence of symmetric coherence peaks ($V_{cp} = \pm 2$ mV) [24]. In addition, the point spectra acquired over the Mn-ML (dashed purple curve) and Mn-DL (dotted green curve) show a similar scenario, even though the reduction of the differential conductance around the Fermi level is less pronounced and the spectra are asymmetric with respect to $U = 0$ mV. Both spectra for the ML and the DL show an increased d$I$/d$U$ in the bias range [–2 mV, 0 mV], which results in the apparent shift of the coherence peak at negative bias towards the Fermi level (particularly visible for the Mn-DL). These observations suggest that: *i)* the Mn thin films are indeed superconducting, due to the proximity effect at the interface with the Nb substrate;

*ii)* YSR bands seem to populate the in-gap local density of states (LDOS) in the Mn, similarly to what is observed for Mn atomic chains on Nb(110) [10], confirming its magnetic nature and motivating the structural and magnetic characterization described below.

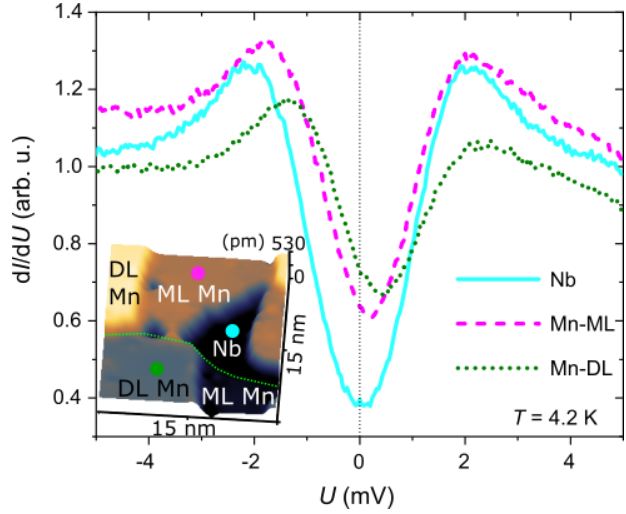

FIG. 1. LDOS on the bare Nb surface (solid cyan), Mn-ML (dashed purple), and Mn-DL (dotted green) characterized via low-bias d$I$/d$U$ spectroscopy (stabilization parameters: $I$ = 1 nA, $U$ = 20 mV; modulation $\Delta U$ = 50 µV) at $T$ = 4.2 K. Each curve is the average of 20 point spectra acquired at the same position over the sample surface. Inset: 3D topographic image showing the locations where the d$I$/d$U$ spectra are acquired. The green dashed line indicates a buried atomic step edge on the Nb(110) surface.

In Fig. 2(a), a large scale topographic image of one of the prepared samples shows an almost fully developed Mn-ML partially covered by small patches of DL, as the result of a step-flow growth. In Fig. 2(b), a line-profile (green curve) shows the apparent height of the deposited thin film over the Nb(110) substrate. Figure 2(c) reports the atomic structure of the Mn-ML, showing a bcc(110) symmetry and interatomic distances along the [001], $a_{Mn}$, and [$\bar{1}$10], $b_{Mn} = \sqrt{2} a_{Mn}$, directions which are in agreement with those of the Nb(110) surface ($a_{Nb}$ = 330 pm, $b_{Nb}$ = 467 pm), demonstrating a pseudomorphic growth. Finally, in Fig. 2(d) the atomic structures of Mn-ML (top-left) and Mn-DL (bottom-right) are directly compared. The pseudomorphic growth is maintained also in the DL patches, which show the same symmetry and same interatomic distances of the ML.

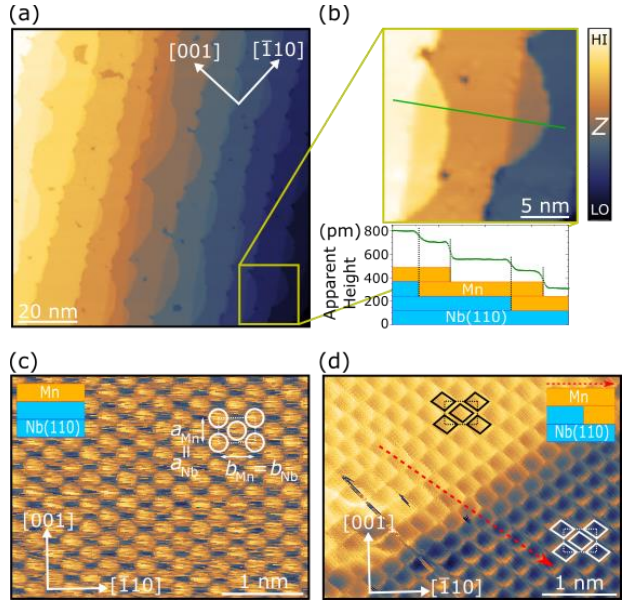

FIG. 2. Structural properties of Mn-ML and Mn-DL on Nb(110) at $T = 4.2$ K. (a) Topographic STM image of 1.2 ML of Mn on the Nb(110) surface ($U = 20$ mV; $I = 2$ nA). (b) Zoom-in from the image in (a), highlighting the presence of a complete ML and patches of DL of Mn, together with a line profile showing the apparent height of the different layers on the sample's surface. (c) Atomically resolved topographic STM image of Mn-ML on Nb(110) ($U = 30$ mV; $I = 2$ nA), showing a bcc(110) surface symmetry, in agreement with the Nb(110) surface. (d) Atomically resolved topographic image (obtained via atomic manipulation imaging mode [32]) of the transition from a Mn-ML to a Mn-DL across a buried step edge on the Nb(110) surface ($U = 3$ mV; $I = 10$ nA). The ML and DL show the same bcc(110) surface symmetry and interatomic distances.

Next, we investigate the magnetic state of the Mn thin films via SP-STM. Figure 3(a) shows a low-bias spin-polarized STM image of the Mn-ML, where the atomic structure and the magnetic contrast are simultaneously resolved. The magnetic contrast in Fig. 3(a) reveals the presence of a spin texture which is consistent with a $c(2\times2)$ antiferromagnetic (AFM) ground state. Due to the symmetry of the bcc(110) Mn surface, in the SP-STM images the $c(2\times2)$ AFM state appears as a row-like pattern [33], where spin-up and spin-down ferromagnetic rows along the [001] direction alternate with each other along the [$\bar{1}$10] direction. These experimental results are in agreement with our DFT calculations, which predict, for the Mn-ML, a $c$(2x2) AFM ground state stabilized by an AFM nearest-neighbor (NN) exchange coupling along the [$\bar{1}$11] direction and reinforced by a FM next-nearest-neighbor (NNN) exchange coupling along the [001] direction (see Fig. 3(b)). Furthermore, an out-of-plane magnetic easy axis is predicted for the Mn-ML, which would establish an AFM ground state as the one shown in the sketch in Fig. 3(c). The spin orientation in the Mn-ML is experimentally verified by imaging its magnetic state with a soft magnetic tip, whose magnetic moment at its apex, $\boldsymbol{m}_{\text{tip}}$, could be easily reoriented via an external magnetic field. Figures 3(d) and 3(e) report the magnetic contrast observed in the Mn-ML via a soft magnetic tip while an out-of-plane magnetic field $B_z = +1$ T and $B_z = -1$ T is applied, respectively. The two images show the same row-like pattern but with an inverted contrast, which is highlighted by the computed *difference* (d) - (e) image presented in Fig. 3(f), where a corrugation of about ±4 pm is observed due to the magnetic contrast contained in the height signal ($z$) of the acquired STM images. All the experimental evidence confirms an AFM ground state with an out-of-plane easy axis, in full agreement with the DFT calculations.

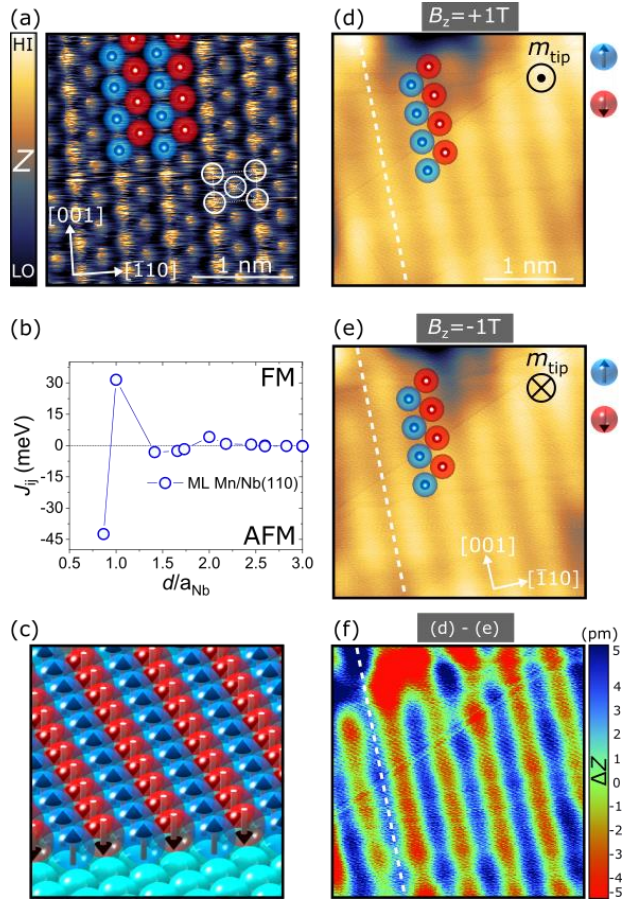

FIG. 3. Magnetism of Mn-ML on Nb(110) at $T = 4.2$ K. (a) SP-STM image of Mn-ML on Nb(110) ($U = 30$ mV; $I = 2$ nA), revealing the presence of the $c(2\times2)$ AFM ground state. (b) Calculated Heisenberg exchange energies, $J_{ij}$, as a function of the inter-atomic distance, $d$, predicting an AFM coupling between NN atoms along $[\bar{1}11]$ and a FM coupling between NNN atoms along [001]. $a_{Nb}$ is the bulk Nb lattice constant. (c) Sketch of the $c(2\times2)$ AFM ground state of the Mn-ML on Nb(110). Blue and red spheres indicate spin-up and spin-down atoms, respectively. (d)(e) SP-STM image of Mn-ML on Nb(110), obtained with a soft magnetic tip while applying a +1 T/-1 T out-of-plane magnetic field ($U = 20$ mV; $I = 5$ nA). (f) Computed *difference* image [(d)-(e)], revealing the out-of-plane magnetic contrast in the Mn-ML.

Finally, we investigate also the magnetic state of the Mn-DL islands and its relation to the spin texture of the adjacent ML areas. In order to access the magnetic state of both the ML and the DL in the same scan, we acquire spin-resolved d$I$/d$U$ maps at $U = +1.9$ V. Figure 4(a) shows a topographic STM image of the sample's surface where both Mn-ML and Mn-DL are present. The spin-resolved d$I$/d$U$ map of the sample's area in the indicated square frame is presented in Fig. 4(b), showing the AFM pattern in both the Mn-ML (pink) and the Mn-DL (green) areas. The dashed black arrow shows how the ferromagnetic row on the ML continues over the DL island situated on the lower atomic terrace, indicating a continuity in the spin orientation. Because we observe this correlation without exemption we conclude that also the Mn-DL has an out-of-plane easy axis. The spin-resolved d$I$/d$U$ map in Fig. 4(b) is acquired with a bulk Cr tip, in the presence of a small out-of-plane field of +0.5 T. In order to test the robustness of the AFM state to high external fields, d$I$/d$U$ maps with $B_z$ = +9 T are acquired as well. Fig. 4(c) and 4(d) show the observed magnetic state in the presence of the high field, demonstrating the immunity of the AFM state of the Mn thin film to the external field. Furthermore, Fig. 4(d) provides additional information concerning the connection of the magnetic state of the Mn-DL to the Mn-ML on the same Nb terrace. As shown by the dashed black arrow oriented along the [001] crystallographic direction, the magnetic contrast is inverted between the DL and the ML on the same atomic terrace. This suggests the presence of an inter-layer NN AFM spin alignment. The outcome of DFT calculations for the exchange interactions in the Mn-DL, reported in Fig. 4(e), are fully in agreement with our experimental observations. Indeed, a particularly strong NN AFM exchange coupling is predicted to dominate the inter-layer magnetic coupling (green symbols + dashed line). In addition, both the bottom (black symbols + solid line) and the top (gold symbols + dotted line) atomic layer are predicted to be characterized by an intra-layer AFM coupling between NN atoms and a weaker FM coupling between NNN atoms, as for the ML case. As a result, the predicted magnetic state is a Type I antiferromagnetic ground state, with the nearest neighbors pointing antiparallel inside the layers as well as between the layers, as schematically shown in Fig. 4(f).

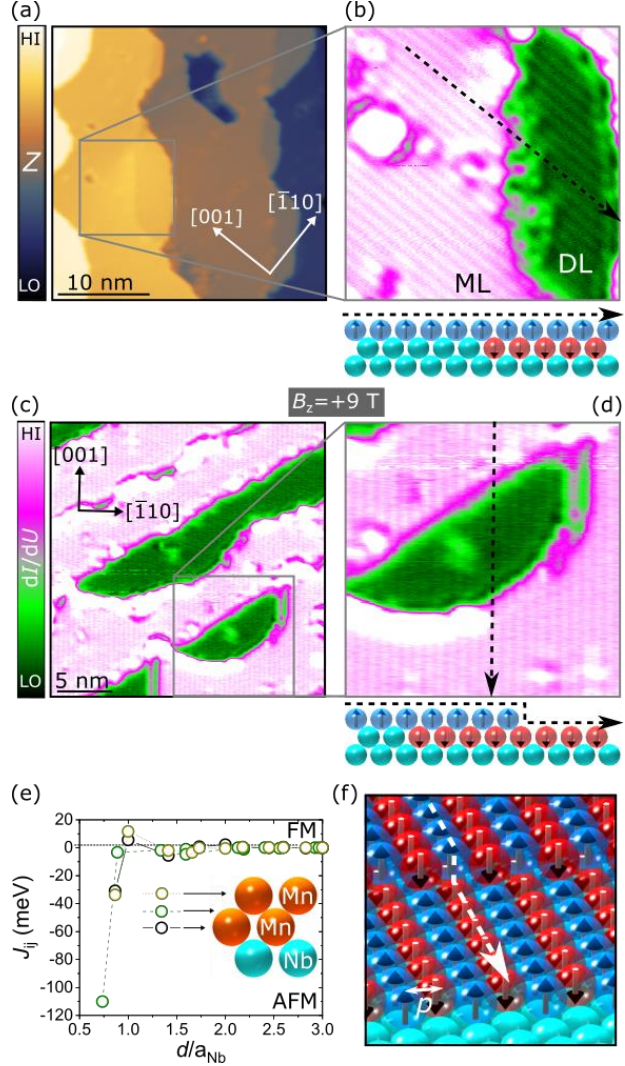

FIG. 4 Magnetism of Mn-ML and Mn-DL on Nb(110) at $T = 4.2$ K. (a) Topographic STM image and (b) spin-resolved differential conductance d$I$/d$U$ map of Mn-ML and Mn-DL on Nb(110). The spin-resolved d$I$/d$U$ map reveals the presence of a $c(2\times2)$ AFM state in both the ML and the DL of Mn, with the magnetic contrast conserved across the ML/DL boundary. (c) Spin-resolved d$I$/d$U$ map of Mn/Nb(110) with an applied +9 T out-of-plane magnetic field, showing an unperturbed AFM state. (d) Zoomed-in d$I$/d$U$ map confirming the conservation of the magnetic contrast going from the ML to the DL across the atomic step of the buried Nb surface and showing the inversion of the magnetic contrast between the top and the bottom atomic layer in the Mn-DL. Imaging parameters: $U = 1.9$ V; $\Delta U = 20$ mV; $I = 5$ nA. Images acquired with a bulk Cr tip. (e) Calculated Heisenberg exchange energies for the Mn-DL as a function of the inter-atomic distance, $d$, predicting a strong inter-layer AFM coupling between NN atoms, which drives the inversion of the AFM spin texture between the bottom and the top atomic layer. (f) Sketch of the AFM state of the bottom and top atomic layer in the Mn-DL on Nb(110). Blue and red spheres indicate spin-up and spin-down atoms, respectively. The white dashed arrow follows the inverted spin state in the top and bottom atomic layer of the Mn-DL, due to a lateral shift of the c(2x2) AFM state by $p = {b_{Mn}}/{2}$ in the $[\bar{1}10]$ crystallographic direction.

It is worth pointing out that the DFT calculations predict a weak in-plane magnetic anisotropy for the extended Mn-DL, with a spin alignment along the $[\bar{1}10]$ direction preferred over the out-of-plane [110] direction by 0.14 meV/atom; while the Mn-ML prefers the out-of-plane spin orientation over the in-plane orientation by 0.54 meV/atom. Accordingly, we should observe a 90° spin reorientation from the [110] to the $[\bar{1}10]$ direction, mostly localized in the Mn-DL due to the almost 4 times larger magnetic anisotropy in the ML and a comparable exchange stiffness. However, we do not have any indication of in-plane magnetization in the experimentally observed DLs. With a bulk Cr tip this transition could be detected in two ways: *i)* the strength of the magnetic contrast, which is proportional to the scalar product between the arbitrarily oriented magnetic moment of the tip, ***m***$_{tip}$, and the local magnetization of the Mn thin film, ***m***$_{Mn}$, would be observed to change between the edge and the center of the DL; *ii)* the magnetic contrast in the ML and the DL would appear either in-phase or inverted, with equal probability over the sample surface due to the degeneracy of the two in-plane magnetic orientations along the easy axis of the DL. However, in none of the acquired spin-resolved images we observe any of those scenarios, supporting the case where an out-of-plane spin texture is present in both layers. We can understand our observations as a finite-size effect. The lateral size of the observed DL islands is smaller or comparable to the calculated domain wall width parameter $\lambda = \sqrt{J_{eff}/K_{eff}}$ = 4.5 nm ($J_{eff}$ and $K_{eff}$ being the calculated effective exchange stiffness and magnetocrystalline anisotropy, respectively), which does not allow for a complete spin reorientation in the middle of the DL to take place.

Finally, it is worth commenting on the coexistence of antiferromagnetism and proximity-induced superconductivity. The antiferromagnetic order of the Mn thin films could play a crucial role for the establishment of the observed proximity-induced superconductivity in our two-dimensional magnet. An AFM order is in principle compatible with the presence of singlet Cooper pairs [34]. Such a scenario was also discussed for the superconducting FeTe monolayer on $Bi_2Te_3$ [35], where the superconducting state was found to coexist with a bi-collinear AFM order. S. Manna and colleagues [35] attributed such coexistence to the fact that the size of the Cooper pairs was larger than the periodicity of the AFM order in their FeTe monolayer, allowing the antiparallel alignment between the spins in the Cooper pairs to survive without phase separation. Even though the superconductivity in the FeTe monolayer is not induced by proximity, the coexistence of superconductivity and antiferromagnetism is analogous to the case observed in our hybrid system, and the same argument could be used in this case, where the size of the Cooper pairs is expected to be comparable to the coherence length in the Nb substrate, $\xi \sim 40$ nm [36], and the periodicity of the AFM order in the Mn film is below the nm. In contrast, a FM state would have had a negative effect on the induced superconductivity, potentially suppressing it by breaking the singlet pairs leaking into the magnetic layer. This clearly establishes Mn thin films on Nb as a promising hybrid system for the study of novel electronic and spin-transport effects in a superconducting antiferromagnet [15,16], and the present atomic scale characterization of the magnetic state of superconducting Mn layers on Nb(110) substrates provides a solid basis for the understanding of potential emergent effects applicable in superconducting antiferromagnetic spintronics.

**(Acknowledgements)** R.L.C. and R.W. acknowledge financial support by the European Union via a Marie Skłodowska-Curie Fellowship (Grant No. 748006 - SKDWONTRACK). R.L.C., A.K., M.B., and K.v.B. acknowledge financial support by the Deutsche Forschungsgemeinschaft (DFG, German Research Foundation) – Project No. 459025680; Project No. 408119516; Project No. 418425860. R.W. acknowledges financial support by the European Union via the ERC Advanced Grant ADMIRE. K.P., L.R. and L.Sz. acknowledge financial support by the National Research, Development, and Innovation Office (NRDI) of Hungary under Project Nos. FK124100 and K131938. K.P. and L.Sz. acknowledge support by the Ministry of Innovation and Technology and the NRDI Office within the Quantum Information National Laboratory of Hungary. K.P. acknowledges the János Bolyai Research Scholarship of the Hungarian Academy of Sciences. The authors would like to thank J. Wiebe, L. Schneider, P. Beck, T. Hänke and E. Mascot for the fruitful discussions.

\* These authors contributed equally.
§ Corresponding Authors

rloconte.magnetism@gmail.com; mbazarni@physnet.uni-hamburg.de